\documentclass[pre,aps,nofootinbib,onecolumn,showpacs]{revtex4}
\usepackage{epsfig}
\usepackage{graphicx}
\usepackage{color}
\usepackage{amssymb,amsmath}
\begin{document}


\title{Griffiths phases in the contact process on complex networks}

\author{G\'eza \'Odor (1), R\'obert Juh\'asz (2), Claudio Castellano (3),
Miguel A. Mu\~noz (4)}
\affiliation{(1) Res. Inst, for Techn. Physics and Materials Science, H-1525
Budapest, P.O.Box 49, Hungary \\
(2) Research Institute for Solid State Physics and Optics, H-1525
Budapest, P.O.Box 49, Hungary \\
(3) Istituto dei Sistemi Complessi (CNR-ISC) UOS Sapienza and 
Dipartimento di Fisica,  ``Sapienza''Universit\'a di Roma, 
P.le Aldo Moro 2, I-00185 Roma, Italy \\
(4) Institute \textit{Carlos I} for Theoretical and Computational
Physics, Universidad de Granada, 18071 Granada, Spain.}  

\begin{abstract} 
Dynamical processes occurring on top of complex networks have become an
exciting area of research. Quenched disorder plays a relevant role in general
dynamical processes and phase transitions, but the effect of {\it topological
quenched disorder} on the dynamics of complex networks has not been
systematically studied so far. Here, we provide heuristic and numerical
analyses of the contact process defined on some complex networks with
topological disorder. We report on Griffiths phases and other rare region
effects, leading rather generically to anomalously slow relaxation in
generalized small-world networks. In particular, it is illustrated that
Griffiths phases can emerge as the consequence of pure topological
heterogeneity if the topological dimension of the network is finite.
\end{abstract}

\maketitle

\section{Introduction}

The study of complex networks has been flourishing since the introduction of a
simple model describing the emergence of scaling in random
networks \cite{BA99}. Multidisciplinary applications involve, for example, the
World Wide Web, various biological, sociological and technological networks.
These are, in many cases, scale-free networks (for recent reviews
see \cite{dorog},\cite{Barab}).  Other families of complex network models are
those composed of a $d$-dimensional regular lattice and additional long
edges \cite{kleinberg}.  These arise e.g. in sociophysics \cite{Barab}, in the
context of conductive properties of linear polymers with cross-links that
connect remote monomers \cite{cc}, in public traffic systems \cite{opttrans},
in the case of nanowires \cite{HBU09}, and in many other examples.  In
general, a pair of nodes separated by the distance $l$ is connected by an edge
with the asymptotic probability for large $l$:
\begin{equation}\label{pdis}
P(l) = \beta l^{-s} \ .
\end{equation}
In the special case $s=0$, the edges exist with a
length-independent probability, as in small world networks, therefore we call
the nets with generic $s$ generalized small world networks (GSW).  If $s\ge
2$, they are characterized by a finite topological dimension $D$,
i.e. $N(l) \sim l^D$, where $N(l)$ is the number of nodes within a distance
$l$ from a given node.  (Note that "small-world" property means that the
number of neighbors of a node increases exponentially with the distance from
it, i.e. formally $D = \infty$).

Dynamical processes defined on networks are of recent interest, for example in
the context of optimization dynamics of spreading or transport
processes \cite{PSV}.  Different dynamical processes, defined on regular
lattices, often exhibit scaling behavior, that can be classified into basic
universality classes \cite{odor}. Instead, the existence of scaling
universality classes on network architectures is not clearly established. In
many cases, the presence of short average distances induces mean-field type of
(fast) dynamics.  It is also well known that structural heterogeneity may lead
to more complex behavior; for example, on scale-free networks, characterized
by a power-law degree distribution $P(k)\propto k^{-\gamma}$
\cite{BA99}, critical exponents may vary with the 
degree exponent $\gamma$ \cite{CR05,dorog}.  In related 'annealed
networks', where the links change rapidly, $\gamma$-dependent critical
exponents have also been reported \cite{BCR08,LHJNP09}. Real networks
are, however, 'quenched' in many cases, i.e. the topology
changes slowly with respect to the dynamical process evolving on them.

In the context of the statistical physics of models defined on
Euclidean lattices, it is well known that quenched randomness can
generate ``rare-region effects'' in the so called {\it Griffiths
phase} (GP), where, in many cases, algebraic scaling is
observed (as opposed to pure systems in which power-laws are
observed only at critical points) with scaling exponents changing
continuously with
the control parameter \cite{Griffiths,GPm,QCP,Vojta,Amir}.  Furthermore, at the
phase transition the evolution becomes logarithmically slow, and it is
controlled by 'activated scaling'.  A nontrivial question can be
posed: Under which conditions do rare-region effects and GPs occur in
network systems?  Can they emerge out of topological disorder alone 
(i.e. without disorder in the transition rates)?  These problems have not been
systematically studied so far. At first sight one might guess that in
network models the dynamics should be very fast, due to the strong
inter-connectedness, i.e. information can propagate very efficiently
throughout the system, resulting in mean-field like, exponential
relaxation.  Recently, however, generic slow dynamics has been
reported to occur in various network models; its origin has been
attributed to heterogeneity \cite{JMT10} or to local bursty activity
patterns \cite{CT-epl,K10}.  To tackle this type of problems, here we
present a study of the contact process (CP) \cite{ContactProcess}
(i.e. the simplest possible model for epidemic spreading, or
information flow) on different GSW networks, including regular
networks and non-regular ones.
%%%%%%%%%%%%%%%%%%%%%%%%%%%%%%%%%%%%%%%%%%%%%%%%%%%%%%%%%%%%%%%%%%%%%%%%%
\section{Numerical analyses}
%%%%%%%%%%%%%%%%%%%%%%%%%%%%%%%%%%%%%%%%%%%%%%%%%%%%%%%%%%%%%%%%%%%%%%%%%

We study the CP on random networks composed of a $d$-dimensional
lattice and a set of long edges with unbounded length. Any pair of
nodes, separated by the distance $l$, is connected by an edge with
some $l$-dependent probability, Eq.(\ref{pdis})
\cite{kleinberg,bb,juhasz}.
An intriguing feature of these type of graphs for $d=1$ is that in the
marginal case ($s=2$) intrinsic properties show power-law behavior and the
corresponding exponents vary continuously with the prefactor $\beta$.  Indeed,
the topological dimension $D$ of such networks has been conjectured to depend
on $\beta$ \cite{bb}.  Instead, for larger values of $s$ long-ranged links are
irrelevant, and $D(\beta)=1$, while for $s<2$ the topological dimension
diverges.  It has been claimed in a recent paper \cite{letter} that if
$D(\beta)$ is finite, Griffiths phases and similar rare-region effects can
appear.

We have considered the CP \cite{ContactProcess}, in which each
infected (active) sites is healed at rate $1$, whereas each of its
nearest-neighbor sites is infected at rate $\lambda /k$ (where $k$ is
the degree of the site) so that the total infection rate is $\lambda$.  In
numerical simulations we update active sites randomly and increment
the time by $1/N_a$, where $N_a$ is the number of active sites. In
this way, all active sites are updated on average once every time
unit.  For a critical infection rate $\lambda_c$ we find a
phase transition from the active to an absorbing phase \cite{AS},
with vanishing density of infected sites.

We have performed numerical simulations initiated either from a fully active
state or from a single active seed. In the former case, the density $\rho(t)$
of active sites has been measured.

%%%%%%%%%%%%%%%%%%%%%%%%%%%%%%%%%%%%%%%%%%%%%%%%%%%%%%%%%%%%%%%%%%%%%%%%%%%%%%%%
\subsection{Non-regular random networks}
%%%%%%%%%%%%%%%%%%%%%%%%%%%%%%%%%%%%%%%%%%%%%%%%%%%%%%%%%%%%%%%%%%%%%%%%%%%%%%%

\begin{figure}[h]
$\begin{array}{c@{\hspace{1cm}}c}
\multicolumn{1}{l} {\mbox{\bf }} &
        \multicolumn{1}{l}{\mbox{\bf }} \\ [-0.53cm]
\epsfxsize=2.5in
\epsffile{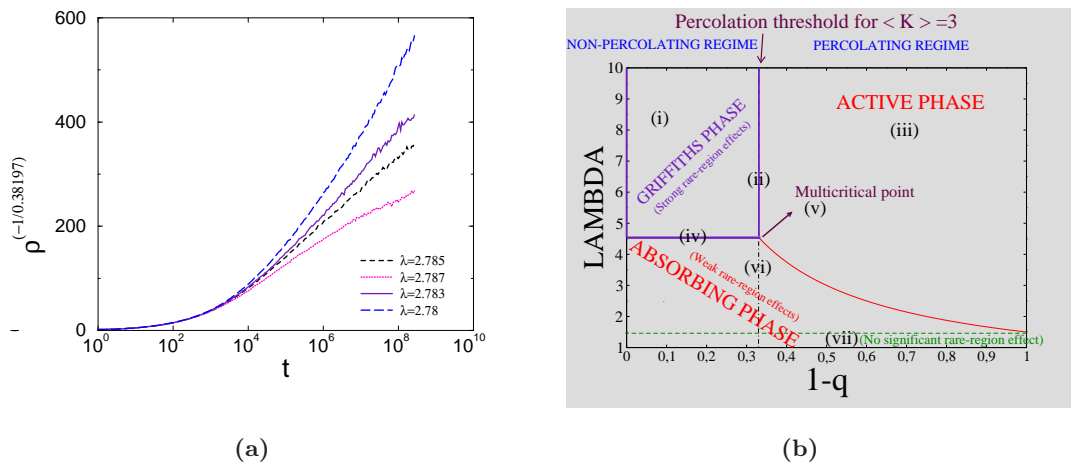} 
&
        \epsfxsize=2.5in
        \epsffile{Fig2.eps} \\ [0.4cm]
\mbox{\bf (a)} & \mbox{\bf (b) }
\end{array}$
\caption{\label{s3fig} Time-dependence of the density in the $s=3$, $\beta=2$
random network. Inset: local slopes of the same (lower curves), and local
slopes of the survival probability in the 1d QCP simulations (dashed
curves). Slow convergence of the effective exponents can be observed.  
(b) Phase diagram of QCP on the ER graph for $r=0$.  }
\end{figure}

The precise definition of the networks we consider is as follows\cite{an,bb}.
We start with $N$ nodes, numbered as $1,2,\dots,N$ and define the distance
between node $i$ and $j$ as $l=\min (|i-j|,N-|i-j|)$.  We connect any pair of
sites separated by a distance $l=1$ (i.e. neighboring sites on the ring) with
probability $1$, and pairs with $l>1$ with a probability $P(l)=1-\exp (-\beta
l^{-s})$.  This implies $P(l) = \beta l^{-s}$ at large distances.  In the case
$s>2$, long-range links do not change the topological dimension, which remains
unity, and hence the critical behavior is expected to be similar to that of
the one-dimensional CP with {\it disordered transition rates}.  In the latter
model, the critical dynamics is logarithmically slow, for instance, starting
from a fully active state, the density decays asymptotically as $\rho \propto
(\ln(t))^{-\tilde\alpha}$ with $\tilde\alpha=0.38197$ \cite{Vojta}.  For $s=3$
(and $\beta=2$) we have run extremely long ($t_{max}=2^{28}$ MCs) simulations
on networks with number of nodes $N=10^5$.  Besides observing a GP, we have
found that data are compatible with the above form of logarithmic dependence
on time, and have located the critical point at $\lambda_c=2.783(1)$.  As
Fig.~\ref{s3fig}(a) shows the assumption on activated scaling with
$\tilde\alpha = 0.38197$ is satisfactory, although only after an extremely
long crossover time to be discussed later \cite{tobepub}.

For the marginal case $s=2$, where the topological dimension is a continuous
function of $\beta$, numerical simulations indicate qualitatively different
scenarios of the phase transition for different values of $\beta$.  If $\beta$
is small enough, GP is observed and the scaling at the transition is of
logarithmic form. However, when $\beta$ is large enough, the GP seems to be
lacking and the critical dynamics follows a power-law.  The latter type of
behavior is observed also for $s<2$, where formally $D=\infty$.

An ``annealed'' counterpart of CP on the above random networks is the 1d CP
with L\'evy flight distributed activation probabilities $P(r) \propto
r^{-d-\sigma}$. In this model the dynamical exponents are known to depend
continuously on $\sigma$ \cite{Hinlrev}.  The estimated dynamical exponent of
the CP on the above random networks in case of power-law critical behavior is
found to be compatible with the dynamical exponent of the L\'evy-flight CP
with $d=1$.  For further details the reader is referred
to \cite{letter,tobepub}.

%%%%%%%%%%%%%%%%%%%%%%%%%%%%%%%%%%%%%%%%%%%%%%%%%%%%%%%%%%%%%%%%%%%%%%%%%
\subsection{3-regular (or cubic) random networks}
%%%%%%%%%%%%%%%%%%%%%%%%%%%%%%%%%%%%%%%%%%%%%%%%%%%%%%%%%%%%%%%%%%%%%%%%%

In the networks studied so far the degree of nodes was random.  In the
following we consider networks with a ``weaker'' topological disorder, in the
sense that the degree of nodes is constant ($3$).  Such random networks with
nodes of degree $3$ can be constructed in the following way \cite{juhasz}.  A
one-dimensional periodic lattice with $N$ sites is given, where the degree of
all sites is initially $2$. Sites of degree $2$ are called ``free sites''.
Let us assume that $N$ is even and $k$ is a fixed positive integer.  A pair of
free sites is selected such that the number of free sites between them is
$k-1$ (the number of non-free nodes can take any value) and this pair is then
connected by a link. That means, for $k=1$, neighboring free sites are
connected, for $k=2$ next-to-neighboring ones, etc. This step, which raises to
$3$ the degree of two free sites, is then iterated until $2(k-1)$ free sites
are left. These are then paired in an arbitrary way, which does not affect the
properties of the network in the limit $N\to\infty$.  In the resulting
network, all sites are of degree $3$, and one can show that the probability of
edges is given by Eq. (\ref{pdis}) with $s=2$.  For certain networks of this
type it has been demonstrated that the long-ranged connections result in a
finite topological dimension which is less than one \cite{juhasz}.
Furthermore, in the case of aperiodic networks, the critical exponents of the
CP depend on the underlying aperiodic modulation \cite{JO09}. In the following, we concentrate on random
networks in which the pairs to be connected are selected randomly (with equal
probability) in the above procedure for a fixed $k$. In this case, one can
show that the prefactor in Eq. (\ref{pdis}) is given by $\beta = k/2$.

\begin{figure}[h]
$\begin{array}{c@{\hspace{1cm}}c}
\multicolumn{1}{l} {\mbox{\bf }} &
	\multicolumn{1}{l}{\mbox{\bf }} \\ [-0.53cm]
\epsfxsize=2.5in
\epsffile{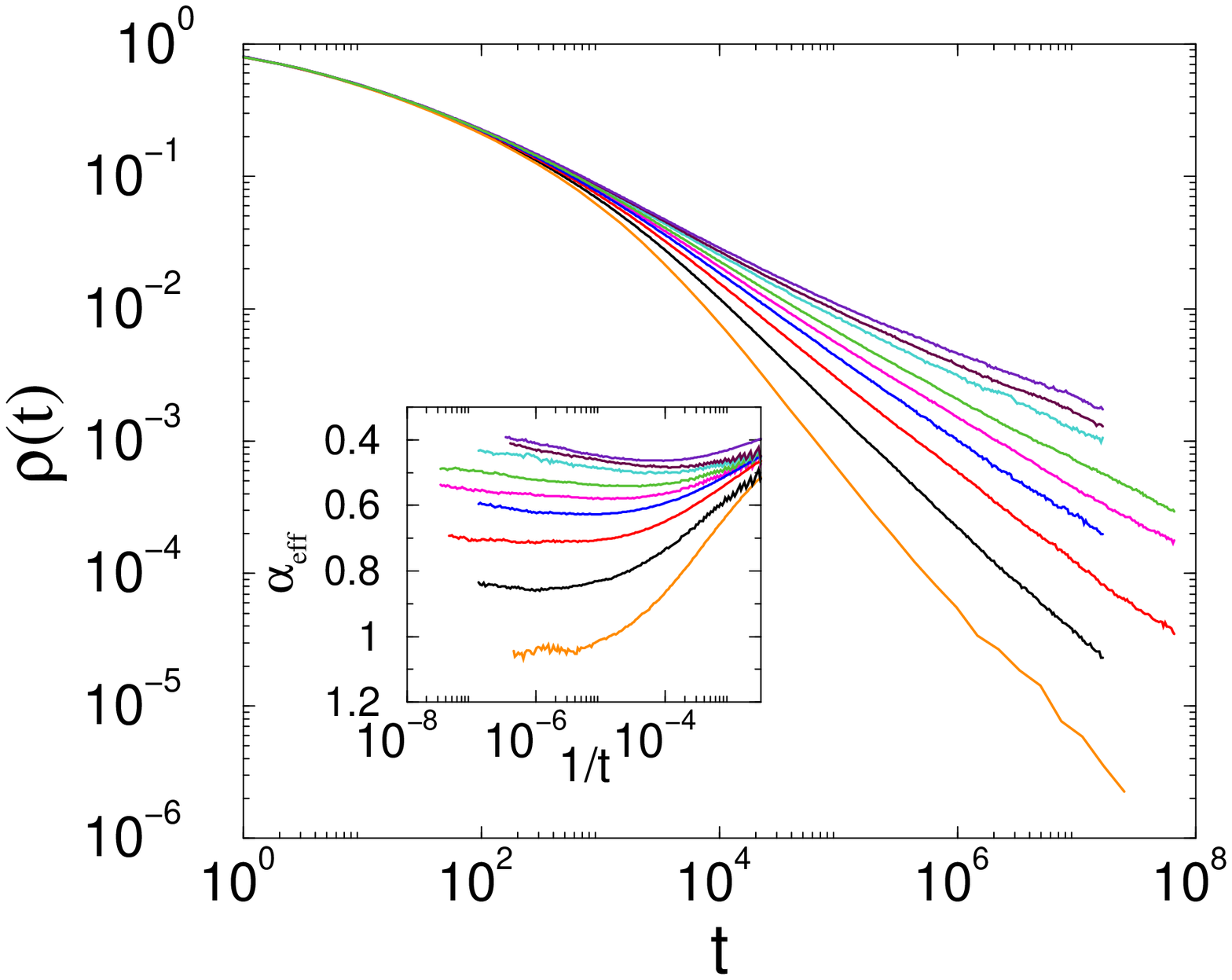} &
	\epsfxsize=2.5in
	\epsffile{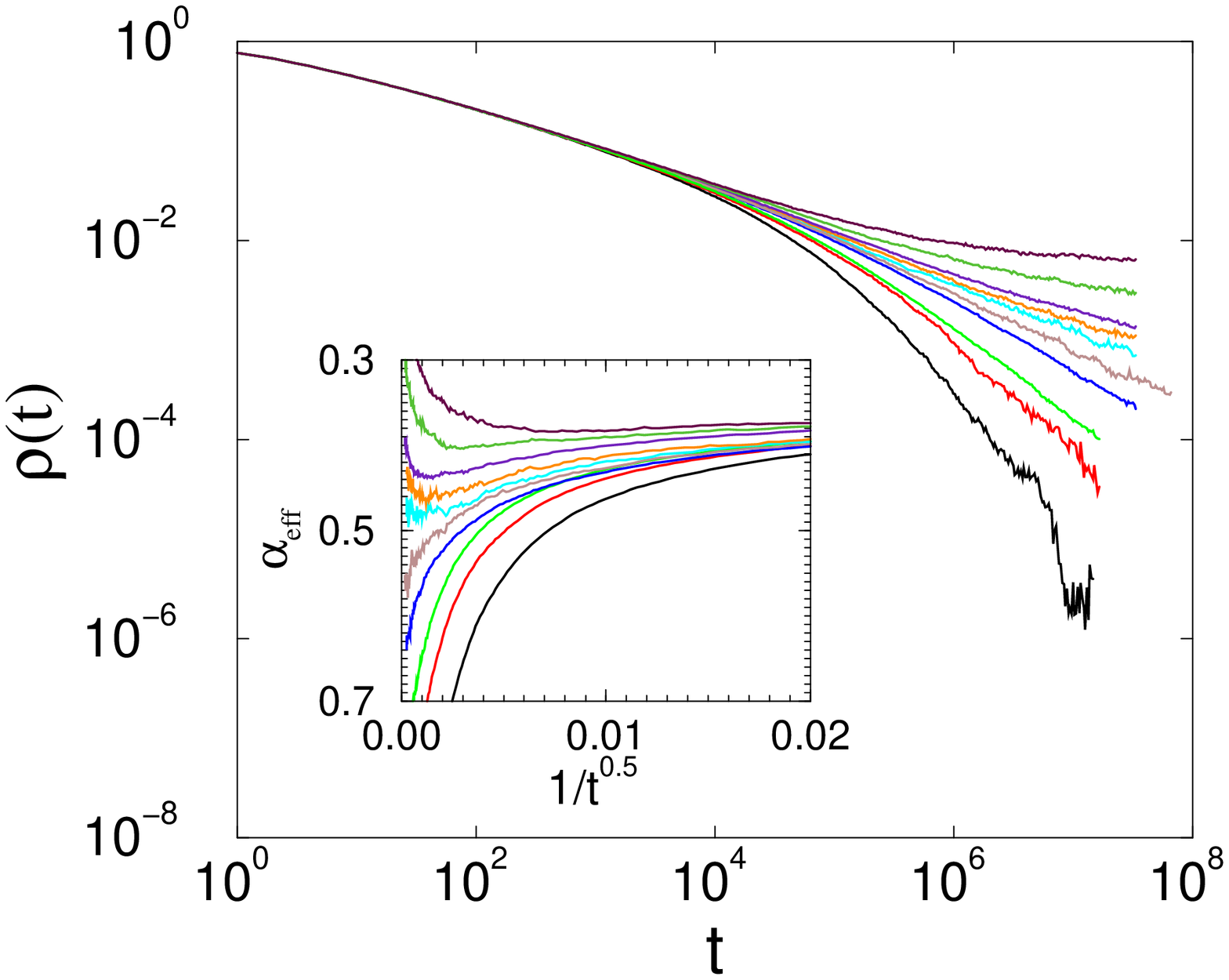} \\ [0.4cm]
\mbox{\bf (a)} & \mbox{\bf (b) }
\end{array}$
\caption{\label{k17} Density decay in 3-regular, random networks.
(a) GP in $k=1$ between $\lambda= 2.535$ and $\lambda=2.57$ (from bottom to
top). (b) Critical point for $k=2$ between $\lambda= 2.156$ and
$\lambda=2.1598$ (from bottom to top).  The inset shows the corresponding
local slopes.}
\end{figure}

We have performed numerical simulations for the CP on $k=1$ random networks
with $N=10^7$ nodes.  The averaging was done over $200$ different network
realizations for each $\lambda$ value and the maximum time is $t_{max}\le
2^{26}$.  As Fig.~\ref{k17}(a) shows, power-law decay with continuously
changing exponent emerges for a range of $\lambda$-s.  By analyzing the
effective decay exponents, defined as the local slopes of the density
$\alpha_{\rm eff}(t) = - (\ln \rho(t) - \ln \rho(t')) / (\ln(t)
- \ln(t^{\prime}))$, where $t/t^{\prime}=2$, the curves do not level-off for
large times, instead one can observe a small drift.  One can easily derive
that the functional dependence of the effective exponent $\alpha_{\rm eff}(t)
= \alpha + x/\ln(1/t)$ is related to the scaling correction $\rho(t) =
t^{-\alpha} \ln^x(1/t)$.  By plotting $\alpha_{\rm eff}(t)$ on logarithmic
time scales (see inset of Fig.~\ref{k17}(a)) it turns out that the drift
describes logarithmic corrections to power-laws and that $\alpha$ is a
non-universal quantity which depends on $\lambda$.  The possibility of
logarithmic corrections has already been pointed out in the case of CP with
quenched disorder \cite{QCP}.

In accordance with results on non-regular networks with $s=2$, the phase
transition is qualitatively different for $k=2$ which corresponds to a higher
value of $\beta$.  Here, a standard phase transition seems to appear at
$\lambda_c=2.15835(5)$, with the density decay exponent $\alpha\simeq 0.51(5)$
(Fig.~\ref{k17}(b)).  This value agrees again with that of the L\'evy flight
CP with $d=1$ and $d+\sigma=2$.

%%%%%%%%%%%%%%%%%%%%%%%%%%%%%%%%%%%%%%%%%%%%%%%%%%%%%%%%%%%%%%%%%%%%%%
\section{QCP on Erd\H os-R\'enyi networks}
%%%%%%%%%%%%%%%%%%%%%%%%%%%%%%%%%%%%%%%%%%%%%%%%%%%%%%%%%%%%%%%%%%%%%%

Apart from studying the effect of heterogeneous (i.e. disordered) topologies
on the dynamics of the contact process, we have also scrutinized the behavior
of the disordered contact process on random networks.  In particular, we have
considered CP on ER graphs \cite{ER} with a quenched disordered infection
rate: a fraction $1-q$ of the nodes (type-I) take value $\lambda$ and the
remaining fraction $q$ (type-II nodes) take a reduced value $\lambda r$, with
$0 \le r < 1$.  Pair mean-field approximations \cite{letter,tobepub} lead to
the following critical threshold
\begin{equation}
\lambda_{c}(q) = \frac{\langle k \rangle}{\langle k \rangle-1} ~
\frac{1}{1-q}.
\label{Crit}
\end{equation}
Type-I nodes experience a percolation transition, where the type~I-to-type~I
average degree is $1$, i.e. at $q_{perc}= 1-\langle k \rangle^{-1}$.  For $q
>q_{perc}$ activity cannot be sustained: type-I clusters are finite and
type-II ones do not propagate activity.

Numerical simulations and optimal fluctuation theory for $\langle k \rangle =
3$ ($q_{perc}=2/3$) leads to the complex phase diagram shown on
Fig.~\ref{s3fig}(b).  In agreement with (\ref{Crit}) one finds a critical
active/absorbing phase transition, but below the percolation threshold the
absorbing phase splits into parts.  In particular, for $\lambda \gtrsim 4.5$
(iv) rare, percolating regions may occur, which cause power-law dynamics (i),
i.e. a Griffiths phase, compared to the inactive phase of simple CP, which is
purely exponential. Further details can be found in \cite{letter,tobepub}.

%%%%%%%%%%%%%%%%%%%%%%%%%%%%%%%%%%%%%%%%%%%%%%%%%%%%%%%%%%%%%%%%%%%%%%
\section{Conclusions}
%%%%%%%%%%%%%%%%%%%%%%%%%%%%%%%%%%%%%%%%%%%%%%%%%%%%%%%%%%%%%%%%%%%%%%

We have illustrated that, besides quenched site disorder, topological disorder
by itself can result in slow dynamics and non-universality, at least for the
contact process. We expect these results to have a broad spectrum of
implications for propagation phenomena and other dynamical process taking
place on networks, and to be relevant for the analysis of both models and
empirical data \cite{JMT10}.  We have claimed in a different
publication \cite{letter} that having finite topological dimension is a
necessary condition for the occurrence of slow, GP type of evolution in
complex networks, at least for CP type of dynamics.  Investigation of other
factors promises to be an interesting, open field of research.

\begin{acknowledgments}
This work was supported by HPC-EUROPA2 pr.228398,
HUNGRID and Hungarian OTKA (T77629,K75324), J. de Andaluc{\'\i}a
P09-FQM4682 and MICINN--FEDER project FIS2009--08451.
\end{acknowledgments}

\end{document}